\documentclass[10pt,reqno]{amsart}
\usepackage[centertags]{amsmath}

\theoremstyle{plain} 

\newtheorem*{conj}{Conjecture}
\newtheorem{thm}{Theorem}

\theoremstyle{definition}

\theoremstyle{remark}
\newtheorem{rem}{Remark}  

\numberwithin{equation}{section}

\DeclareMathOperator{\diag}{diag} 
\DeclareMathOperator{\rank}{rank} \DeclareMathOperator{\id}{Id}

\DeclareMathSymbol{\R}{\mathalpha}{AMSb}{"52}
\DeclareMathSymbol{\C}{\mathalpha}{AMSb}{"43}

\newcommand{\set}[1]{\left\{#1\right\}}

\newcommand{\pd}{\partial \,}
\newcommand{\fpd}[2]{\frac{\pd #1}{\pd #2}}
\newcommand{\spd}[2]{ #2 \frac{\!\pd #1}{\!\pd #2}}

\newcommand{\ra}{\rightarrow}
\newcommand{\beq}{\begin{equation}}
\newcommand{\eeq}{\end{equation}}
\newcommand{\bd}{\begin{description}}
\newcommand{\ed}{\end{description}}
\newcommand{\beqr}{\begin{eqnarray}}
\newcommand{\eeqr}{\end{eqnarray}}

\begin{document}

\title[Invariants of differential equations]
{Invariants of differential equations defined by vector fields}
\author[J C Ndogmo]{ J C  Ndogmo}

\address{
PO Box  2446\\
Bellville 7535\\
South Africa.}
\email{ndogmoj@yahoo.com}
\begin{abstract}
We determine the most general group of equivalence transformations
for a family of differential equations defined by an arbitrary
vector field on a manifold. We also find all invariants and
differential invariants for this group up to the second order. A
result on the characterization of classes of these equations by the
invariant functions is also given.
\end{abstract}

%
\maketitle

\section{Introduction}
In Lie theory, the invariance of functions and other objects under a
transformation group $\mathcal{G}$ acting on an $n$-dimensional
manifold $V$ is usually characterized by the vanishing of functions
under some vector fields generating the group action, and this
vanishing is represented by a system of partial differential
equations of the form
\begin{equation} \label{eq:det0}  \sum_i A_i(X)\, \pd_ {\!x^i}  F(X)=0
 \end{equation}
 where $X=(x^1,..., x^n)$ is a local coordinates system on $V$
and where $F$ is the unknown
 function. The importance of linear partial differential equations of
 the form \eqref{eq:det0} usually referred to as determining equations
 for the invariant objects cannot be overstated. Indeed, they characterize
 invariant equations as well as  their invariant solutions, and they
 have a similar importance in the study of Lie algebras and in
 representation theory. In physics, invariant operators of dynamical
 groups characterize specific properties of physical systems and
 provide mass formulaes and energy spectra \cite{gell-m, engf}.
 Invariants of physical symmetry groups also provide quantum numbers
 useful in the classification of elementary particles \cite{wig}.
 It would therefore be desirable to consider
 the group of equivalence transformations of equations of the form
 \eqref{eq:det0}  and to determined all functions invariant under
 this group. Such functions are simply called invariants of the
 differential equation \eqref{eq:det0}.\par
  A method for the determination of invariants of linear and
  nonlinear equations build on an idea suggested by Lie
  himself \cite{lie1} was developed in \cite{theo}. The method
  is based on the fact that the invariant functions for the infinite group of
  equivalence transformations of a given system of equations are
  precisely the invariants of the system of differential  equations. That is, these
  functions are invariant under the group of
  transformations that confine the system of equations to a prescribed family
   of equations. The method also yields singular invariant
   equations, and has been
   used to complete the problem of determination of the Laplace
   invariants in \cite{laplace}, and  in
   \cite{lin2} to characterize linearizable second order
   {\sc ode}'s.\par
   In the present paper we find the most general group $G$ of equivalence
   transformations leaving unchanged,  except
   for its coefficients $A_i$, an equation of the form
\begin{equation} \label{eq:det1}  \sum_{i=1}^n A_i(X)\, \pd_ {\!x^i}  U=0
\end{equation}
where $(x^1, \dots, x^n, U)\in \R^n \times \R=M,$ and where we
assume that none of the coefficients $A_i$ for $i=1, \dots, n$
vanishes identically. We then find the invariants and differential
invariants up to the second order for this group and for an
arbitrary number $n$ of independent variables in the equation. We
first treat with more details the cases $n=2,3$ before giving some
generalizations of the results. Next, by investigating the
regularity of  the action of $G$ on $M,$ we show how the invariants
found can be used to characterize families of equations of the form
\eqref{eq:det1}.


\section{The group of equivalence transformations }
\label{s:group}
  Owing to the linearity of equation \eqref{eq:det1}, any invertible change
of the dependent variable $U$ and the independent variables $(x^1,
\dots, x^n)=X$ that preserves the form of the equation should be of
the form
\begin{subequations}\label{eq:chg1}
\begin{align}
X &= \psi(Y)     \label{eq:ind1}\\
U &= H (Y) V(Y), \qquad H(Y)\, \neq 0 \label{eq:dep1}
\end{align}
\end{subequations}
where $Y=(y^1, \dots, y^n)$ is the new set of independent variables,
$V$ is the new dependent variable and $H$ is an arbitrary function.

\begin{thm}\label{th:equivg}
The most general group $G$ of equivalence transformations of
equation \eqref{eq:det1} consists of the set of all invertible
changes of variables  of the form
\begin{subequations}\label{eq:chg2}
\begin{align}
x^i &= \psi^i(Y)\equiv  \psi^i(y^i), \qquad \text{for $i=1, \dots, n$}    \label{eq:ind2}\\
U &=  V \label{eq:dep2}.
\end{align}
\end{subequations}
That is, each $\psi^i (Y)$ is a function a the single variable
$y^i,$ and $G$ does not involve a change of the dependent variable.
\end{thm}

\begin{proof}
Under the general change of variables \eqref{eq:ind1}, and by
setting $\phi= \psi^{-1},$ equation \eqref{eq:det1} takes the form
\begin{subequations}\label{eq:chg3}
\begin{align}
 \sum_j & B_j (Y) \pd_ {\!y^j} (U)=0   \label{eq:chg3.1}\\[-6mm]
\intertext{ where} B_j(Y) &= \sum_i^n A_i (\psi(Y))\frac{\pd
\phi^j}{\pd x^i}(\psi (Y))= \sum_i^n A_i(X) \frac{\pd \phi^j}{\pd
x^i} (X) \label{eq:chg3.2}
\end{align}
\end{subequations}
Equation \eqref{eq:chg3.1} together with the expression of $U$ given
by \eqref{eq:dep1} shows that none of the coefficients $B_j$ should
vanish identically. However,  the expression of $B_j$ in
\eqref{eq:chg3.2} shows that if $\phi^j (X)$ depends on more than
one of the variables $x^i, \text{ for $i=1, \dots, n$}$ it can be
chosen as an invariant of an appropriate vector field, and so that
$B_j=0.$ Hence $\phi^j(X) \equiv \phi^j(x^{qj}),$ for some $qj \in
\set{1,\dots, n}$ and because of the invertibility of $\phi,$
$\phi^j$ must be a nonconstant map and  all the variables $x^{qj}$
must be distinct for $j=1, \dots,n.$ This implies in particular that
$x^i= \psi^i (y^{ki})$ must also be a nonconstant function of a
single variable. If we let $\sigma$ be the permutation that maps the
ordered set $\set{y^1, \dots, y^n}$ onto the ordered set
$\set{y^{k1}, \dots, y^{kn}},$ then the $i$th component of $\psi
\circ \sigma^{-1}$ depends exactly on $y^i$ alone. On account of the
arbitrariness of $\psi,$ we may replace $\psi,$ by $\psi \circ
\sigma^{-1},$ and thus that we may always assume that $x^i= \psi^i
(y^i),$ and equivalently $y^i= \phi^i(x^i).$ This reduces the
expression for $B_j(Y)$ in \eqref{eq:chg3.2} to the form

\begin{equation} \label{eq:Bj} B_j=  A_j(\psi(Y))\, \frac{\pd \phi^j (x^j)}{\pd
x^j} =  \frac{A_j(\psi(Y))}{\psi^{j\, \prime}(y^j)} \neq 0,
\end{equation}
where $\psi^{j\, \prime}= \pd \psi^j /\pd y^j.$ Substituting
\eqref{eq:dep1} into \eqref{eq:chg3.1} and expanding, equation
\eqref{eq:det1} takes the form

\begin{align}\label{eq:chg4}
 \sum_j & H\, B_j \, \pd_ {\!y^j}V +V\left( \sum_j B_j \, \pd_ {\!y^j}\, H \right)
 =0.
\end{align}
The fact that the coefficient of $V$ appearing in \eqref{eq:chg4}
must identically vanish and the arbitrariness of the $n$
coefficients $A_j$ in the expression of $B_j$ in \eqref{eq:Bj} show
that $\pd_ {\!y^j}\, H (Y)=0,$ for all $j=1, \dots,n.$ Thus $H(Y)
\neq 0$ is a constant function and without loss of generality we may
assume that $H=1.$ This last equality transforms equation
\eqref{eq:chg4} to the form

\begin{equation} \label{eq:chgfn}\sum_j  B_j (Y) \pd_ {\!y^j}V(Y) =0, \end{equation}
which is of the prescribed form. This completes the proof of the
theorem.
\end{proof}
\begin{rem}
 It should also be noted that under the general change of
variables \eqref{eq:chg1}, it is always possible, by the well-known
result on the rectification of vector fields, to put \eqref{eq:det1}
in the form
$$\pd_{\!y^1} (H V)=0, \quad \text{ that is, }\quad (\pd_{\!y^1} H) V+ H(\pd_{\!y^1} V)=0  $$
Thus if we allow some of the coefficients $A_i$ to vanish, then the
only additional condition to be imposed on the change of variables
\eqref{eq:chg1} would be $\pd_{\! y^1} H=0,$ and all equations of
the form \eqref{eq:det1} would be equivalent. There are clearly no
invariant functions or invariant equations of any order in such
case.
\end{rem}
We now move on to determine the infinitesimal generators of the
group $G.$ As already noted, equation \eqref{eq:ind2} implies that
$y^i= \phi^i(x^i), \text{ for $i=1, \dots, n$},$ and this shows that
the  infinitesimal transformation of \eqref{eq:chg2} has the form
\begin{equation} \label{eq:infi1} y^i \approx x^i + \epsilon \xi^i(x^i), \qquad
V \approx U,
\end{equation}
where the functions $\xi^i$ are also arbitrary, due to the
arbitrariness of the functions $\psi^i.$ The first prolongation of
this transformation has the form
\begin{equation}
\pd_{\! y^i} V \approx \pd_{\! x^i} U+ \epsilon (- \xi^{i \, \prime}
\pd_{\! x^i} U ),
\end{equation}
which implies that
\begin{equation}\label{eq:infi2}
\pd_{\! x^i} U \approx \pd_{\! y^i} V + \epsilon (\xi^{i \, \prime}
\pd_{\! y^i} V,
 )
\end{equation}
where $\pd_ {\!x}$ is the differential operator $\pd / \pd x,$ for
any variable $x.$ A substitution of equation \eqref{eq:infi2} into
the original equation \eqref{eq:det1} yields the infinitesimal
transformation of that equation in the form
\begin{equation}
\sum_i^n (A_i + \epsilon A_i\,  \xi^{i \, \prime} )\pd_{\! y^i}V =0.
\end{equation}

This shows that the infinitesimal transformation ${\tilde A}$ of the
coefficient $A_i$ is given by
$$
{\tilde A_i} \approx A_i + \epsilon A_i \xi^{i \, \prime}.
$$
The infinitesimal generators of the equivalence transformation $G$
therefore has the form
\begin{equation}\label{eq:v}
\mathcal{V} = \sum_i^n \xi^i \pd_{\! x^i} + \sum_i^n A_i\,
 \xi^{i \, \prime} \pd_{\! A_i}
\end{equation}


\section{Zeroth-order invariants}
\label{s:inv1}
We would like to first recall very briefly certain elementary facts
about the invariant functions of a given transformation group.
Suppose that the infinitesimal generators of an $r$-parameters group
of transformations $G$ acting on the $\mathcal{Q}$-dimensional
manifold $M$ are of the form
\begin{equation} \label{eq:vk's}
\mathcal{V}_k = \sum_j \xi^{kj}\pd_{\! x^j}, \qquad \text{ for $k=1,
\dots, r.$}
\end{equation}
The invariant functions and invariant equations of $G$ are
determined by 
\begin{subequations} \label{eq:invcdt}
\begin{align}
\mathcal{V}_k\, (F)=0 \label{eq:detinv}\\
\mathcal{V}_k\,( F) \Bigl\vert_{F=0}=0 \label{eq:detinveq}
\end{align}
\end{subequations}
respectively, $\text{ for $k=1, \dots, r$}.$ The number
 of fundamental invariants of $G$ does not exceed
$\mathcal{Q}-\tau$, where $\tau$ is the rank of the matrix $\left(
\xi^{kj}\right)_{k,j}$ of coefficients of the $r$ operators
$\mathcal{V}_k.$ Each of these functions naturally gives rise to an
invariant equation. Invariant equations $F=0,$ where $F$ is not an
invariant function, and obtained by imposing the additional
condition $\tau< \mathcal{Q}$ to the second equation of
\eqref{eq:detinveq} are often referred to as singular invariant
equations. Using a Lie linearization test, such equations were
recently shown \cite{lin2} to characterize all linearizable second
order ordinary differential equations. \par
   When some of the independent variables $x^j$ in the expression of
the  $\mathcal{V}_k$ can be taken as depend variables for other
objects such as a differential equation, the generators
$\mathcal{V}_k$ can be extended to involve higher order derivatives
of the dependent variables. If $\mathcal{V}$ is a given
infinitesimal generator of $G,$ then we shall often use the same
symbol $\mathcal{V}$ to represent both $\mathcal{V}$ and its $m$-th
prolongation $\mathcal{V}^{(m)}$. Similarly, the $m$th jet space of
$M$ will often be denoted simply by $M.$\par
Since the general change of variables \eqref{eq:chg2} is merely a
change of the independent variables and does not involve the
dependent variable $U,$ this variable is trivially an invariant for
$G.$ We shall therefore ignore this variable in our search for the
invariant functions of $G$ whose general form for the zeroth-order
operator \eqref{eq:v} is $F(x^1, \dots,x^n, A_1,\dots, A_n ).$
\begin{thm}
The group of equivalence transformations $G$ of \eqref{eq:det1} has
neither invariant functions  nor invariant equations.
\end{thm}
\begin{proof}
 Rewriting the generic generator
$\mathcal V$ in \eqref{eq:v} as a linear combination of the
arbitrary functions $\xi^i$ and their derivatives gives
$$
\mathcal{V}= \sum_i^n \xi^i (\pd_{\! x^i}) + \sum_i^n
\xi^{i\,\prime} (A_i\, \pd_{\! A_i}),
$$
and this proves the first part of the theorem at once, on account of
 the arbitrariness of the functions $\xi^i.$ To show that $G$
has no invariant equation, we use an elementary technique similar to
that used in \cite{lin2}. Suppose that $F(x^1, \dots,x^n, A_1,\dots,
A_n )=0 $ is a nontrivial invariant equation for $G,$ and so it
explicitly involves at least one of the variables, say $x^1,$ in the
set $\set{x^1, \dots, x^n, A_1, \dots, A_n}.$ Then solving the
equation for $x^1$ reduces it to the equivalent form $x^1=
K(x^2,\dots, x^n, A_1, \dots, A_n).$ The arbitrariness of the
functions $\xi^i$ and their derivatives implies again that we must
have in particular
$$ \pd_{\! x^1} \left(x^1- K\right)\Bigl \vert_{\,x^1=K}=0.
$$
But this last condition cannot hold because $\pd_{\! x^1} \left(x^1-
K\right)=1,$ and this completes the proof of the theorem.
\end{proof}


\section{First-order differential invariants}
The first prolongation $\mathcal{V}^{(1)}$ of the infinitesimal
generator \eqref{eq:v} of $G$ has the form
\begin{subequations}
\begin{align} \label{eq:plg1}
\mathcal{V}^{(1)}&= \mathcal{V}+ \sum_i^n A_i \xi^{i\, \prime
\prime} \frac{\pd}{\pd A_{ii}} + \sum_i^n \sum_{j\neq i} A_{ji}
\left(
 \xi^{j\, \prime}- \xi^{i\, \prime} \right) \frac{\pd}{\pd
 A_{ji}}\\
\intertext{where we have used the notation}
A_{ij}&= \frac{\pd A_i}{\pd x^j}, \qquad \text{ for $i, j\in \set{1,
\dots, n}.$}
\end{align}
\end{subequations}
In terms of the linear combination of the arbitrary functions
$\xi^i$ and their derivatives, this expression takes the form
\begin{equation}\label{eq:xiplg1}
\begin{split}
\mathcal{V}^{(1)}&= \sum_i^n \xi^i \frac{\pd}{\pd x^i}
 + \sum_i^n
\xi^{i\, \prime \prime} \frac{\pd}{\pd A_{ii}}\\
& + \sum_i^n \xi^{i\, \prime} \left[\frac{\pd}{\pd A_i} +
\sum_{j\neq i} \left(A_{ij} \frac{\pd}{\pd
 A_{ij}} - A_{ji} \frac{\pd}{\pd
 A_{ji}}\right)\right].
\end{split}
\end{equation}
Equation \eqref{eq:xiplg1} clearly shows that any first-order
differential invariant of $G$ should be independent of all the
independent variables $x^i,$ as well as the variables $A_{ii}.$ This
last condition reduces $\mathcal{V}^{(1)}$ to the form
\begin{subequations} \label{eq:xip1}
\begin{align}
\mathcal{V}^{(1)} &= \sum_i^n  \xi^{i\, \prime}\,
\mathcal{V}_{\xi^{i\, \prime}}\\[-2.5mm]
\intertext{where}
\mathcal{V}_{\xi^{i\, \prime}}&=  A_i \pd_{\! A_i} + \sum_{j\neq i}
\left(A_{ij} \frac{\pd}{\pd
 A_{ij}} - A_{ji} \frac{\pd}{\pd
 A_{ji}}\right).
\end{align}
\end{subequations}
As we are considering the arbitrary functions  $A_{i}$ in
\eqref{eq:det1} in their most general form, we may assume that the
functions $A_{ij}= \pd_{\! x^j} A_i$ do not vanish identically. It
then follows from \eqref{eq:xip1} that the generators of the first
prolongation of $G$ depend on $\mathcal{Q}=n^2$ independent
variables. If however we assume that exactly $p$ of the functions
$A_{ij}$ vanish identically, then this number $p$ is invariant under
the action of $G$ and $\mathcal{Q}= n^2-p.$
\begin{thm}\label{th:vxipl1}
Consider the $n$ operators $\mathcal{V}_{\xi^{i\, \prime}}$ of
\eqref{eq:xip1}.
\begin{enumerate}
\item[(a)] The rank of the coefficients matrix $\mathcal{M}$ of the operator $\mathcal{V}_{\xi^{i\,
\prime}}$ is $n,$ which is maximal.
\item[(b)] The $\mathcal{V}_{\xi^{i\, \prime}}$ form an
$n$-dimensional commutative Lie algebra.
\item[(c)] The  number fundamental first-order
differential invariants of the group $G$ of equivalence
transformations of equation \eqref{eq:det1} is $n(n-1)$.
\end{enumerate}
\end{thm}
\begin{proof}
In any coordinate system of the form $\set{A_1, \dots, A_n, \dots}$
on the extended jet space on which the first prolongation of $G$
operates, equation \eqref{eq:xip1} shows that the first $n$ columns
of $\mathcal{M}$  is represented by the matrix $\diag\set{A_1,
\dots, A_n}$ which has rank $n,$ owing to the fact that none of the
coefficients $A_i$ is zero, and this proves part (a) and shows that
the $n$ vectors $\mathcal{V}_{\xi^{i\, \prime}}$ are linearly
independent. For part (b), if  for any $k \in \set{1,\dots, n}$ we
write
\begin{align*}
\mathcal{V}_{\xi^{k\, \prime}}&=  A_k \pd_{\! A_k} + \sum_{q\neq
 k} \left(A_{kq} \frac{\pd}{\pd
 A_{kq}} - A_{qk} \frac{\pd}{\pd
 A_{qk}}\right)
\end{align*}
 then we readily see that the
commutator $\left[\mathcal{V}_{\xi^{i\, \prime}},
\mathcal{V}_{\xi^{k\, \prime}} \right]$ is a linear combination of
the identically vanishing commutators
\begin{align*}
\left[ A_{ij}\pd_{\! A_{ij}} - A_{ji}\pd_{\! A_{ji}} ,A_{kq}\pd_{\!
A_{kq}} -A_{qk}\pd_{\! A_{qk}}
 \right],& \qquad \left[A_i\pd_{\! A_i},  A_k\pd_{\! A_k} \right]\\
\left[ A_i\pd_{\! A_i}, A_{kq}\pd_{\! A_{kq}} -A_{qk}\pd_{\! A_{qk}}
\right],& \qquad  \left[ A_{ij}\pd_{\! A_{ij}} - A_{ji}\pd_{\!
A_{ji}} , A_{k}\pd_{\! A_{k}} \right],
\end{align*}
 This fact together with part (a) proves (b). Since the number of
independent variables involved in the complete system of $n$
operators $\mathcal{V}_{\xi^{i\, \prime}}$ is $n^2,$ the number of
their functionally independent invariants is precisely
$n^2-\rank(\mathcal{M})$, which is  $n(n-1).$
\end{proof}
The most practical way to find the $n^2-n$ first order differential
invariants of $G$ would be to compute these invariants for low
dimensions of $M,$ i.e. for $n=2,3$ and then make use of the
symmetry inherent in equation \eqref{eq:det1} to find the invariants
in the general case.\par
For $n=2$ and $n=3$ we write equation \eqref{eq:det1} in the form
\begin{equation}\label{eq:neq23} a U_x + b U_y =0, \qquad \text{ and }
\quad  a U_x + b U_y + c U_z=0,
\end{equation}
respectively.  In case $n=2,$ the operators $\mathcal{V}_{\xi^{i\,
\prime}}$ are given by
\begin{equation*}
\mathcal{V}_{\xi^{1\, \prime}} =a \pd_a + a_y \pd_{\! a_y} - b_x
\pd_{\! b_x}, \qquad \mathcal{V}_{\xi^{2\, \prime}} = b\pd_b - a_y
\pd_{\! a_y} + b_x \pd_{\! b_x}.
\end{equation*}
Solving the system of equations $\mathcal{V}_{\xi^{i\, \prime}}
(F)=0, \text{ for $i=1,2$}$ by the method of characteristics shows
that $G$ has a fundamental system of invariants consisting of the
two functions
$$ T_{12}=\frac{a_y\, b}{a}, \quad \text{ and } \quad T_{21}=\frac{b_x \,a}
{b}.
$$
 In case $n=3,$ the three operators $\mathcal{V}_{\xi^{i\,
\prime}}$ are given by
\begin{align*}
\mathcal{V}_{\xi^{1\, \prime}}&= a \pd_a + (a_y \pd_{\! a_y} - b_x
\pd_{\! b_x})
                                 + (a_z \pd_{\! a_z} - c_x \pd_{\! c_x}) \\
\mathcal{V}_{\xi^{2\, \prime}}&= b\pd_b + (b_x
\pd_{\! b_x}  - a_y \pd_{\! a_y})+ (b_z \pd_{\! b_z} -  c_y\pd_{\! c_y}) \\
\mathcal{V}_{\xi^{3\, \prime}}&= c \pd_c+ (c_x \pd_{\! c_x} -
a_z\pd_{\! a_z}) + (c_y \pd_{\! c_y} - b_z\pd_{\! b_z})
\end{align*}
and the corresponding set of six invariants is found to be
\begin{align*}
T_{12}&=\frac{a_y\, b}{a}, \quad  T_{21}=\frac{b_x \,a}
{b}, \quad T_{13}=\frac{a_z\, c}{a}\\
T_{31}&=\frac{c_x\, a}{c}, \quad  T_{23}=\frac{b_z \,c} {b}, \quad
T_{32}=\frac{c_y\, b}{c}.
\end{align*}
The form of the invariants found for $n=2,3$ together with part (c)
of Theorem~ \ref{th:vxipl1} asserting that the number of invariants
in the general case is $n(n-1),$ which is $2 \binom{n}{2},$ suggest
that all invariants can be found by associating with each subset of
two elements of the set of $n$ coefficients of the differential
equation a pair of invariants according to a very simple rule.
\begin{thm}
The $n(n-1)$ fundamental invariants $T_{ij}$ of the group $G$ of
equivalence transformations of equation \eqref{eq:det1} are given by
\begin{equation} \label{eq:allinv1}
T_{ij}= \frac{A_{ij} A_j}{A_i}, \quad \text{ for $i\neq j, \quad $
where}\quad  A_{ij}= \frac{\pd A_i}{\pd x^j},
\end{equation}
and where $A_i$ and $A_j$ run over the set  coefficients of the
equation.
\end{thm}
\begin{proof}
It is easily verified that the identity $\mathcal{V}_{\xi^{i\,
\prime}} (T_{kq})=0$ holds for all $i=1,\dots,n$ and for all $k\neq
q.$ Next, the functions $A_{ij}$ for $i,j \in \set{1, \dots,n}$ are
functionally independent by assumption, and each $T_{ij}$ depends on
exactly one of them.
\end{proof}
Note that if we restrict the action of $G$ to a sub-family of
equations of the form  \eqref{eq:det1} for which exactly $p$ of the
function $A_{ij}$ vanish identically, then the maximal number of
functionally independent first order differential invariants is
$n(n-1)-p.$
\section{Second-order differential invariants}
The second prolongation of the generator \eqref{eq:v} of $G$ has the
form
\begin{subequations} \label{eq:pl2v}
\begin{align}
\mathcal{V}^{(2)} =\; \mathcal{V} &+ \sum_j^n (A_j ) \xi^{j\, \prime
\prime} \pd_{\! A_{jj}} + \left(A_j \xi^{j\, \prime \prime}+ A_{jj}
\xi^{j\, \prime \prime} - A_{jjj} \xi^{j\, \prime }
\right)\pd_{\! A_{jjj}} \notag \\
\qquad&+ \sum_{i\neq j} A_{ji} (\xi^{j\, \prime } -\xi^{i\, \prime
}) \pd_{\! A_{ji}}+ 2 \left( A_{ji}\xi^{j\, \prime \prime} -
A_{jji}\xi^{i\, \prime } \right)\pd_{\! A_{jji}} \notag\\
\qquad&+ \left[(\xi^{j\, \prime} - 2 \xi^{i\, \prime }) A_{jii}
-A_{ji}
\xi^{i\, \prime \prime} \right]\pd_{\! A_{jii}}\notag \\
\qquad&+2 \sum_{\substack{i,k \,\neq j\\i<k} } (\xi^{j\, \prime }
-\xi^{i\, \prime }-\xi^{k\, \prime }) A_{jik}\, \pd_{\!
A_{jik}},\\[-2.5mm]
\intertext{where as usual}
& A_{ji} = \frac{\pd A_j}{\pd x^i}, \qquad A_{jik}= \frac{\pd
A_j}{\pd x^i \pd x^k},\quad  \text{etc.} \label{eq:ajs}
\end{align}
\end{subequations}
Rewriting this expression as a linear combination of the arbitrary
functions $\xi^{i}$ and their derivatives shows that any invariant
function should be independent from the independent variables and
from variables of the form $A_{iii}$ for $i=1,\dots,n.$ This reduces
the expression of $\mathcal{V}^{(2)}$ to the form
\begin{subequations} \label{eq:xipl2}
\begin{align}
\mathcal{V}^{(2)} =\,  &\sum_i^n \xi^{i\, \prime }
\mathcal{V}_{\xi^{i\, \prime }} + \xi^{i\, \prime \prime}
\mathcal{V}_{\xi^{i\, \prime \prime}}\\[-2.5mm]
\intertext{where}
\mathcal{V}_{\xi^{i\, \prime }}=\, &A_i \pd_{\! A_i} + \sum_{j \neq
i} \left(  A_{ij} \pd_{\! A_{ij}} -A_{ji}
\pd_{\! A_{ji}} + A_{ijj} \pd_{\! A_{ijj}}  - 2 \sum_k^n A_{jik} \pd_{\! A_{jik}} \right) \notag \\
\parbox{0.3in}{$\qquad \qquad$}&+ 2 \sum_{\substack{j,k \,\neq i\\j<k} }
A_{ijk} \pd_{\! A_{ijk}} \label{eq:vxi2p}\\
\mathcal{V}_{\xi^{i\, \prime \prime}} =\, &A_i\, \pd_{\! A_{ii}} +
\sum_{j\neq i} \left(  2 A_{ij}\, \pd _{\! A_{iij}}- A_{ji} \,
\pd_{\! A_{jii}}\label{eq:vxi2pp} \right).
\end{align}
\end{subequations}
It readily follows from equations \eqref{eq:xipl2} that the second
order differential invariants of $G$ depend in general on $n+
n\binom{n+2}{2}- 2n$ variables, that is on $n^2(3+n)/2$ variables.
This is the dimension of the subspace of the extended jet space
$M^{(2)}$ of $M$ on which the second prologation of $G$ acts.
\begin{thm}\label{th:vxipl2}
The set of operators  $\set{\mathcal{V}_{\xi^{i\, \prime}}}_{i=1}^n$
and $\set{\mathcal{V}_{\xi^{i\, \prime \prime}}}_{i=1}^n$ given in
\eqref{eq:xipl2} each generate an $n$ dimensional commutative Lie
algebra.
\end{thm}
\begin{proof}
Thanks to the  term $A_i \pd_{\! A_i}$ appearing in the expression
of each generator $\mathcal{V}_{\xi^{i\, \prime}}$ as the only term
involving $\pd_{\! A_{i}},$ the coefficients matrix of these
operators admits a submatrix of the form $\diag\set{A_1, \dots,
A_n},$ which is clearly of rank $n$, showing that the
$\mathcal{V}_{\xi^{i\, \prime}}$ generate an $n$-dimensional space.
Similarly, as the term $A_i \pd_{\! A_{ii}}$ appears in the same
manner in the expression of each generator $\mathcal{V}_{\xi^{i\,
\prime \prime}}$, the set $\set{\mathcal{V}_{\xi^{i\, \prime
\prime}}}_{i=1}^n$ also generate an $n$ dimensional space. For each
pair $\set{i,k},$ it is easy to see as in the proof of Theorem
\ref{th:vxipl1} that each of the commutators $[\mathcal{V}_{\xi^{i\,
\prime}}, \mathcal{V}_{\xi^{k\, \prime}}]$ and
$[\mathcal{V}_{\xi^{i\, \prime \prime}}, \mathcal{V}_{\xi^{k\,
\prime \prime}}]$ is a linear combination of identically vanishing
commutators. This completes the proof of the theorem.
\end{proof}
 If we denote by $\xi^{i\, (j)}$ the $j$th derivative of $\xi^i,$ then Theorem
\ref{th:vxipl2} asserts that for $j$ fixed, the $\mathcal{V}_{\xi^{i
(j)}}$'s form a commutative Lie algebra for $j=1,2.$ However, the
set of all operator $\mathcal{V}_{\xi^{i (j)}}$ for $i=1, \dots, n$
and $j=1,2$ that determine the second order differential invariants
of $G$ do not form a Lie algebra in general when they are considered
together, as this easily appears from the low dimensional cases.\par
  Indeed, if for $n=2,3$ we rewrite equation \eqref{eq:det1} as in
\eqref{eq:neq23}, then for $n=2,$ we have
\begin{align*}
\mathcal{V}_{\xi^{1 \, \prime}} &= a\fpd{ }{a} + a_y \fpd{}{a_y}+
a_{yy}\fpd{}{a_{yy}} -b_x\fpd{}{b_x} -2b_{xx}\fpd{}{b_{xx}}
-2b_{xy}\fpd{}{b_{xy}}\\
\mathcal{V}_{\xi^{2 \, \prime}}&= b\fpd{ }{b}  -a_y\fpd{ }{a_y}
-2a_{xy}\fpd{ }{a_{xy}} - 2a_{yy}\pd_{\! a_{yy}} + b_x\fpd{ }{b_x}+
b_{xx}\fpd{}{b_{xx}}\\
\mathcal{V}_{\xi^{1 \, \prime \prime}}&= a\fpd{ }{a_x} + 2a_y\fpd{ }{a_{xy}} -b_x\fpd{ }{b_{xx}}\\
\mathcal{V}_{\xi^{2 \, \prime \prime}}&=  -a_y\fpd{ }{a_{yy}}
+b\fpd{ }{b_{y}} + 2 b_x\fpd{ }{b_{xy}}.
\end{align*}
In this case we have
$$
[\mathcal{V}_{\xi^{1 \, \prime}} , \mathcal{V}_{\xi^{1 \, \prime
\prime}}]= \mathcal{V}_{\xi^{1 \, \prime \prime}}, \quad \text{ and
} \quad [\mathcal{V}_{\xi^{2 \, \prime}} , \mathcal{V}_{\xi^{2 \,
\prime \prime}}]= \mathcal{V}_{\xi^{2 \, \prime \prime}}
$$
However, the  span of $\set{{V}_{\xi^{1 \, \prime}}, {V}_{\xi^{2 \,
\prime}}, \mathcal{V}_{\xi^{1 \, \prime \prime}},\mathcal{V}_{\xi^{2
\, \prime \prime}} }$ does not contain the commutator
$[\mathcal{V}_{\xi^{i \, (j)}} , \mathcal{V}_{\xi^{k \, (p)}}]$ for
any sets $\set{i,k}$ and $\set{j,p}$  of distinct elements. For
instance, we have
$$
[\mathcal{V}_{\xi^{1 \, (2)}} , \mathcal{V}_{\xi^{2 \, (1)}}]= -2
a_y \,\fpd{}{ a_{xy}}.
$$
We have a similar situation in the case of three independent
variables. The operators $\mathcal{V}_{\xi^{i \, (j)}}$ are given in
this case by
\begin{align*}
\mathcal{V}_{\xi^{1 \, \prime }}&=\,  \spd{}{a} +  \spd{}{a_y} +
\spd{}{a_z} +
\spd{}{a_{yy}} + 2\, \spd{}{a_{yz}} + \spd{}{a_{zz}}  \notag \\
&-\spd{}{b_x} - 2\spd{}{b_{xx}} -2 \spd{}{b_{xy}} -2\spd{}{b_{xz}} -\spd{}{c_x}\notag \\
 &-2 \spd{}{c_{xx}} -2\spd{}{c_{xy}}-2 \spd{}{c_{xz}} \\[2mm]
\mathcal{V}_{\xi^{ 2\, \prime }}&=\, \spd{}{b}- \spd{}{a_y} -2 \spd{}{a_{xy}} -2 \spd{}{a_{yy}} -2 \spd{}{a_{yz}}+ \spd{}{b_x} \notag \\
& + \spd{}{b_z} + \spd{}{b_{xx}} +2 \spd{}{b_{xz}} + \spd{}{b_{zz}}- \spd{}{c_y} \notag\\
&-2 \spd{}{c_{xy}} -2 \spd{}{c_{yy}} - 2 \spd{}{c_{yz}}\\[2mm]
\mathcal{V}_{\xi^{3 \, \prime }}&=\,  \spd{}{c} -\spd{}{a_z} -2
\spd{}{a_{xz}}- 2\spd{}{a_{yz}}
-2 \spd{}{a_{zz}} - \spd{}{b_z} \notag\\
&- 2 \spd{}{b_{xz}}-2 \spd{}{b_{yz}} -2 \spd{}{b_{zz}} +
\spd{}{c_x} + \spd{}{c_y}\notag\\
& + \spd{}{c_{xx}} + 2\spd{}{c_{xy}} + \spd{}{c_{yy}}\\[2mm]
\mathcal{V}_{\xi^{1 \, \prime \prime}}&=\, a\fpd{}{a_x} + 2 a_y \fpd{}{a_{xy}} + 2 a_z
\fpd{}{a_{xz}} - b_x \fpd{}{b_{xx}} - c_x \fpd{}{c_{xx}} \\[1mm]
\mathcal{V}_{\xi^{2 \, \prime \prime}}&=\, -a_y \fpd{}{a_{yy}}+ b\fpd{}{b_y} + 2 b_x \fpd{}{b_{xy}} + 2 b_z \fpd{}{b_{yz}}- c_y
\fpd{}{c_{yy}}\\[1mm]
\mathcal{V}_{\xi^{3 \, \prime \prime}}&=\, - a_z \fpd{}{a_{zz}} -
b_z \fpd{}{b_{zz}} + c\fpd{}{c_z}+ 2c_x \fpd{}{c_{xz}} + 2 c_y
\fpd{}{c_{yz}}.
\end{align*}
As in case $n=2,$ we have $[\mathcal{V}_{\xi^{i \, \prime}} ,
\mathcal{V}_{\xi^{i\, \prime \prime}}]= \mathcal{V}_{\xi^{i \,
\prime \prime}}.$  That is, $\set{\mathcal{V}_{\xi^{i \, \prime}} ,
\mathcal{V}_{\xi^{i\, \prime \prime}}}$ spans a solvable Lie algebra
with nilradical $\set{\mathcal{V}_{\xi^{i\, \prime \prime}}},$ for
$i=1,2,3.$ However, here again the span of $\set{
 \mathcal{V}_{\xi^{i \, (j)}} }_{i,j}$ does not contain the commutator
$[\mathcal{V}_{\xi^{i \, (j)}} , \mathcal{V}_{\xi^{k \, (p)}}]$ for
any sets $\set{i,k}$ and $\set{j,p}$  of distinct elements. Indeed,
we have for instance
$$[\mathcal{V}_{\xi^{1 \, (2)}} , \mathcal{V}_{\xi^{3 \, (1)}}]= 2
c_x \fpd{}{ c_{xz}}.$$
There is no guarantee in this case that the number of invariant
attains its maximum which is $\mathcal{Q}-\tau,$ with the usual
notation. More over, they are much more difficult to find using the
method of characteristic. we shall therefore attempt to determine
the invariants of the second prolongation of $G$ using the so-called
method of total derivatives \cite{forsyth, ndog04}.\par
Suppose that we are given a system of equations of the form
\eqref{eq:detinv} where the $\mathcal{V}_k$'s are arbitrary linear
differential operators given as in \eqref{eq:vk's} and depend on a
total of $\mathcal{Q}$ variables. Denote again by $\tau$ the rank of
the coefficients matrix $\left( \xi^{kj}\right)_{k,j},$ and set
$p=\mathcal{Q}-\tau.$ Thus we can solve \eqref{eq:detinv} for $\tau$
of the variables $\pd_{\! x_t}F$ in terms of the remaining $p$
others, and this gives rise to the  Jacobian system
\begin{equation} \label{eq:jacobian}
\Delta_t F \equiv \fpd{F}{x_t} + \sum_{s=1}^p U_{s,t}
\fpd{F}{u_s}=0, \qquad \text{ for $t=1, \dots, \tau$},
\end{equation}
where we have renamed the remaining $p$  variables $x_{\tau +j}$ as
$u_j,$ for $j=1, \dots, p$ and where the $U_{s,t}$'s are functions
depending in general on the $\mathcal{Q}$ vriables $x_1, \dots,
x_{\tau}$ and  $u_1, \dots, u_p.$ In this case the equivalent
adjoint system of total differential equations takes the form
\begin{equation} \label{eq:adjoint}
d u_s= \sum_{t=1}^{\tau} U_{s,t} d x_t, \qquad \text{ for $s=1,
\dots, p.$}
\end{equation}
The equations \eqref{eq:detinv} and \eqref{eq:adjoint} are
equivalent in the sense that they have the same integrals
\cite{forsyth}. We denote by $\mathcal{M}_u$ the coefficients matrix
$\set{U_{s,t}}$ that completely determines the adjoint system
\eqref{eq:adjoint}.\par
For $n=2,$ by permuting  the coordinate system on $M$ so as to have
in  $\diag \set{a, b, a, b}$ as the submatrix corresponding to the
first 4 columns of the coefficients matrix for the system
$\mathcal{S}_{22}=\set{{V}_{\xi^{1 \, \prime}}, {V}_{\xi^{2 \,
\prime}}, \mathcal{V}_{\xi^{1 \, \prime \prime}},\mathcal{V}_{\xi^{2
\, \prime \prime}} },$ we obtain the transposed matrix
$\mathcal{M}_u^T$ of $\mathcal{M}_u$ in the form
%
\begin{equation*}
\mathcal{M}_{u}^T= \begin{pmatrix}
 0                    &  \frac{a_{yy}}{a}    &-\frac{b_x}{a} & \frac{a_y}{a}   &-\frac{ 2 b_{xx}}{a} & -\frac{ 2b_{xy}}{a}\\[1.3mm]
 -\frac{2 a_{xy}}{b}  & -\frac{2 a_{yy}}{b}  &\frac{b_x}{b}  & -\frac{ a_y}{b} & \frac{b_{xx}}{b}    & 0
 \\[1.3mm]
\frac{2 a_y}{a}       & 0                    & 0             & 0               & -\frac{b_x}{a}      & 0
\\[1.3mm]
 0                    & -\frac{a_y}{b}       & 0             & 0               & 0                   & \frac{2b _x}{b}
\end{pmatrix}.
\end{equation*}
The corresponding system \eqref{eq:adjoint} of total differential
equations can be solved using  methods described in \cite{forsyth}.
We  get stuck with a problem of finding some integrating factors
while trying to solve by the method of characteristics the
equivalent system \eqref{eq:detinv} of linear partial differential
equations for the system of operators $\mathcal{S}_{22}.$ However,
we readily get the following set of six functions by solving the
corresponding adjoint system \eqref{eq:adjoint}.
\begin{equation}\label{eq:inv2v2}
\begin{split}
&T_{12}= \frac{a_y\, b}{a}, \quad K_{12}= \frac{a_{yy}\, b}{a_y} +
b_y, \quad  J_{12}= \frac{a_{xy}\, ab}{a_y} - 2 a_x\\
&T_{21}= \frac{b_x\, a}{b}, \quad K_{21}= \frac{b_{xx}\, a}{b_x} +
a_x, \quad  J_{21}= \frac{b_{xy}\, ab}{b_x} - 2 b_y.
\end{split}
\end{equation}
Although their number corresponds to the maximal number of
functionally independent invariants in this case, not all of them
are actually invariants because the system $\mathcal{S}_{22}$ is not
complete. More precisely, only $T_{ij}$ and $K_{ij},$ for $i,j=1,2$
are invariants, and not only the $J_{ij}$'s are not invariants, but
 also the equations $J_{i,j}=0$ are not invariant equations.\par
   Similarly for $n=3,$ the total number $\mathcal{Q}$  of variables defining the
invariants is  $27,$ and we have $\tau=6.$ By permuting again the
coordinate system on $M,$ so as to have $\diag \set{a,b,c, a,b,c}$
as the first six columns of the coefficients matrix for the system
of perators $\mathcal{S}_{32}=\set{{V}_{\xi^{1 \, \prime}},
{V}_{\xi^{2 \, \prime}},{V}_{\xi^{3 \, \prime}}, \mathcal{V}_{\xi^{1
\, \prime \prime}},\mathcal{V}_{\xi^{2 \, \prime
\prime}},\mathcal{V}_{\xi^{3 \, \prime \prime}} },$ we obtain a more
convenient representation of the $21 \times 6$ matrix $\mathcal{M}_u
.$ The transpose $\mathcal{M}_u^T$ of $\mathcal{M}_u$ in which only
its first six columns are represented has the form
\begin{equation*}
\mathcal{M}_{u}^T= \begin{pmatrix}
0&0    &\frac{a_{yy}}{a}&\frac{2 a_{yz}}{a}   &\frac{a_{zz}}{a} &-\frac{b_x}{a}   & \dots \\[1.3mm]
-\frac{2 a_{xy}}{b}&0   &-\frac{2 a_{yy}}{b}  &-\frac{2 a_{yz}}{b}  &0   &\frac{b_x}{b}  & \dots  \\[1.3mm]
0& -\frac{2 a_{xz}}{c}     &0      &-\frac{2 a_{yz}}{c}  &-\frac{2 a_{zz}}{c}     &0  & \dots    \\[1.3mm]
\frac{2 a_y}{a}& \frac{2 a_z}{a}      &0             &0    & 0                 & 0& \dots \\[1.3mm]
0    &0     &-\frac{a_y}{b}             &0 &0 & 0& \dots   \\
   0             &0     &0             &0 &-\frac{a_z}{c} & 0& \dots
\end{pmatrix},
\end{equation*}
where the dots represent the remaining $15$ matrix columns. Solving
the corresponding system \eqref{eq:adjoint} yields the expected
maximal number of 21 functionally independent functions. But since
here again the corresponding system of operators $\mathcal{S}_{32}$
is not complete, only $15$ of them are actually invariants of $G.$
Reverting back to the original notation $A_1=a, A_2=b$ and $A_3=c,$
and using \eqref{eq:ajs}, these $15$ invariants can be written in
the form
\begin{equation}\label{eq:inv3v2}
T_{ij}= \frac{A_{ij}\, A_j}{A_i},\quad K_{ij}= \frac{A_{ijj} \,
A_j}{A_{ij}} + A_{jj}, \quad L_{ijk}= A_{ijk} \left( \frac{A_j
A_k}{A_i}\right),
\end{equation}
where $i,j \in \set{1,2,3},$ with $i\neq j,$ and where
$\set{j,k}=\set{1,2,3}\setminus \set{i}$ for $i=1,2,3.$
We have thus obtained the following result.
\begin{thm}
Let $\mathcal{N}$ be the maximal number of functionally independent
invariants of the second prolongation  of the group of equivalence
transformations of \eqref{eq:det1} in $n$ independent variables.

\begin{enumerate}
\item[(a)] For $n=2,$ $\mathcal{N}=4,$ and the invariants are the
function $T_{ij}$ and $K_{ij}$ of \eqref{eq:inv2v2}.
\item[(b)] For  $n=3,$ $\mathcal{N}=15,$ and the invariants are
the functions $T_{ij}, K_{ij}$ and $L_{ijk}$ given by
\eqref{eq:inv3v2}.
\end{enumerate}
\end{thm}
Contrary to the case of the first prolongation of $G$, a
determination of all invariants of the second prolongation for
larger values of $n$ using only invariants of a lower order of $n$
and a symmetry argument does not seem to be obvious. Indeed, the
equations \eqref{eq:inv2v2} and \eqref{eq:inv3v2} show that for
$n=3,$ the invariants of type $T_{ij}$ and $K_{ij}$ can be simply
derived by symmetry from those for $n=2$ without any further
calculations. However, the invariants of type $L_{ijk}$ in
\eqref{eq:inv3v2} cannot be obtained from \eqref{eq:inv2v2} using
only a symmetry argument. This makes it more difficult to find all
the invariants  for the second prolongation of $G$ when $n \geq 4.$
Nevertheless, we do have the following result which is solely based
on a symmetry argument.
\begin{thm}
For $n\geq 3,$ a fundamental set of invariants of the second
prolongation of $G$ includes all the invariants of type $T_{ij}$,
$K_{ij}$ and $L_{ijk}$ of \eqref{eq:inv3v2}, whose total number is
$n(n^2+n-2)/2.$
\end{thm}
Indeed, this result clearly follows from \eqref{eq:inv3v2} and the
symmetry inherent in \eqref{eq:det1}, by noting that the total
number of the $T_{ij}, K_{ij}$ and $L_{ijk}$ for
 $n\geq 3$ is
\begin{equation} \label{eq:nbnvar1}
2 \binom{n}{2} + 2 \binom{n}{2} + 3 \binom{n}{3}=
\frac{1}{2}\,n(n^2+n-2).
\end{equation}
Although we may not find all the invariants of the second
prolongation of $G$ for larger values of $n >3$ using only symmetry
arguments, it should be possible to predict their number. Denote by
$M_k^{n,j}$ the number of fundamental invariants of the $k$th
prolongation of equation \eqref{eq:det1} (with $n$ independent
variables) involving terms of the form $A_{I},$ where $I$ is an
index of the form $i_1 i_2 \dots i_j$ with distinct $i_k \in \set{1,
\dots, n}$ for $k=1, \dots, j.$  Note that the corresponding type of
functions appears for the first time as invariants of
\eqref{eq:det1} when the number of independent variables is $j,$
where $2\leq j\leq n.$ If we also denote by $M_k^n$ the number of
invariants of the $k$th prolongation for $n$ variables, then a
closer look at equation \eqref{eq:inv3v2} suggests that $A_2^{n,j}=
j \binom{n}{j}$ and $M_2^n= M_1^n+ W_n,$ where
$$ W_n= A_2^{n,2} + A_2^{n,3}+ \dots + A_2^{n,n}= \sum_{j=2}^n j \binom{n}{j}.$$
Using the properties of binomial coefficients, it can be shown that
$\sum_2^n j \binom{n}{j}$ equals $n(2^{n-1}-1).$ Since by Theorem
\ref{th:vxipl1} we have $M_1^n= n(n-1),$ our conjecture follows.
\begin{conj}
For any value $n$ of independent variables in equation
\eqref{eq:det1}, the number $M_2^n$ of functionally independent
invariants of the second prolongation of $G$ is $n(2^{n-1}+n-2).$
\end{conj}
This conjecture says that $M_2^4=40,$ and $M_2^5=95.$ By a result of
Lie (see \cite{laplace}), it is possible to find differential
invariants of $G$ of higher order than $2$ using invariant
differentiation, but we will not discuss that here.

\section{Properties of the invariants}
It follows from Theorem \ref{th:equivg} that every element of the
equivalence transformations group $G$ of equation \eqref{eq:det1}
can be represented by an invertible map $\phi$ of the form
$$ \phi \colon \R^n \ra \R^n \colon X=(x^1, \dots, x^n)
\mapsto Y=\phi(X) \equiv (\phi^1 (x^1), \dots, \phi^n(x^n)).$$
That is, the $i$th component of $\phi$ depends only on the single
variable $x^i.$ In a given coordinate system $X= (x^1, \dots, x^n),$
each equation of the form \eqref{eq:det1} can be represented by the
$n$-tuple $\left(A_i(X)\right)_{i=1}^n$ or just $(A_i(X)).$ It
follows from equation \eqref{eq:Bj} that under the action of $\phi
\in G,$ the differential equation $(A_i(X))$ is mapped to the
differential equation $(B_i(Y))= \phi \cdot (A_i(X))$, where
$$ B_i(Y)= A_i(\phi^{-1} (Y))  \phi^{i\, \prime} (\psi^i(Y)) =
\frac{A_i(\psi(Y))}{\psi^{i\, \prime}(y^i)},$$
and where $\psi= \phi^{-1}.$ If $\theta$ is any other element of
$G,$ and $\id$ is the identity transformation,  it is easy to see
that
\begin{equation} \label{eq:gact}
\id \cdot (A_i(X))= (A_i(X)), \quad  \text{ and } \quad  \theta~
\cdot~ \left( \phi \cdot (A_i(X))\right)= \theta \circ \phi \cdot
(A_i(X)).
\end{equation}
 Thus if we denote by
$E_n$ the variety of all differential equations of the form
\eqref{eq:det1}, then \eqref{eq:gact} shows that the action of $G$
on $M$ induces another group action of $G$ on $E_n.$ This yields a
partition of $E_n$ into  orbits which can be described by the
original action of $G$ on $M.$ For a given element $(A_i(X))$ in
$E_n,$ we set
$$T_{ij}^A = A_{ij} A_j /A_i.$$
\begin{thm}\label{th:orbits}
Suppose that the coefficients $A_i,$ for $n=1, \dots, n$ in equation
\eqref{eq:det1} are non vanishing and that exactly $p$ of the
functions $A_{ij}$ in the expression of the generators
$\mathcal{V}_{\xi^{i\, \prime}}$ of the first prolongation of $G$ in
\eqref{eq:xip1} vanish identically. Then two differential equations
$(A_i (X))$ and $(B_i (X))$ are equivalent, i.e. they belong to the
same orbit of the first prolongation of $G$ if and only if
$$ T_{ij}^A (X)= T_{ij}^B (X) $$
for all of the $n(n-1)-p$ nonzero such functions $T_{ij}^A$ and
$T_{ij}^B.$
\end{thm}
\begin{proof}
If the coefficients $A_i$ of \eqref{eq:det1} are non vanishing, then
$\diag \set{A_1, \dots, A_n}$ has constant rank $n,$  and by the
expression of the generators in \eqref{eq:xip1}, $G$ acts
semi-regularly. More over, the expression of the corresponding
invariant functions in \eqref{eq:allinv1} shows that this action is
regular. Since the invariants of $G$ are actually the invariants of
the induced group action of $G$ on $E_n,$ the result follows from a
theorem of \cite[Theorem 2.34]{olver} stating that for regular group
actions, two points lie in the same orbit if and only if they take
on the same values under all invariant functions.
\end{proof}

\section{Concluding Remarks}
It follows from Theorem \ref{th:orbits} above that all differential
equations  $(A_i(X))$ where \\$A_i=A_i (x^i)$ depends on $x^i$ alone
are equivalent. The expression of the generators in \eqref{eq:xip1}
shows that the action of $G$ restricted to this family of equations
has no invariant of any order.


%
\end{document}